\documentclass[graybox]{svmult}

\usepackage{mathptmx}       
\usepackage{helvet}         
\usepackage{courier}        
\usepackage{type1cm}        
\usepackage{makeidx}         
\usepackage{graphicx}        
\usepackage{multicol}        
\usepackage[bottom]{footmisc}

\makeindex             

\begin{document}

\title*{Solar-like stars observed by Kepler: an incredible adventure}
\author{R.A. Garc\'\i a}
\institute{R.A. Garc\'\i a on behalf of the WG$\#$1 Team \at Laboratoire AIM, CEA/DSM-CNRS-Universit\'e Paris Diderot; IRFU/SAp, Centre de Saclay, 91191, Gif-sur-Yvette, France, \email{rgarcia@cea.fr}
}
%
%
\maketitle

\abstract*{The NASA {\it Kepler mission} --in flight since March 2009-- is producing an enormous number of high-quality continuous light curves. Now, and for the first time ever, we are able to do Òensemble asteroseismologyÓ, i.e., to do an asteroseismic analysis with a statistically significant sub-sample of solar-like stars covering a wide range of stellar characteristics. In the present work, I highlight some of the most recent studies carried out using these data.}

\abstract{The NASA {\it Kepler mission} --in flight since March 2009-- is producing an enormous number of high-quality continuous light curves. Now, and for the first time ever, we are able to do Òensemble asteroseismologyÓ, i.e., to do an asteroseismic analysis with a statistically significant sub-sample of solar-like stars covering a wide range of stellar characteristics. In the present work, I highlight some of the most recent studies carried out using these data.}

\section{Introduction}
\label{sec:Intro}
{\it Kepler} \cite{Bor10} is a planet-hunter mission to look for Earth-size planets around solar-like stars in the habitable zone. The photometric stability and the quiet environment provided by the Earth trailing heliocentric orbit offers a great opportunity to perform asteroseismic observations of thousands of stars (after some specific processing of the light curves \cite{Gar11}) among the 150,000 observed by {\it Kepler}  at a single star field in the Cygnus-Lyra region of the galaxy. {\it Kepler} is planned to work for 3.5 years with a possible extension to 6 years. With its telescope of 0.95-m aperture, {\it Kepler} monitored 100,000 stars at any time with a cadence of 29.42 min (long-cadence measurements). A subsample of 512 stars can be observed at a much faster cadence of 58.85s (short-cadence measurements; Nyquist frequency of $\sim$ 8.5 mHz) allowing for more precise transit timings \cite{Gil10}. This running mode is of particular interest for asteroseismology because it allows us to study stochastically-excited oscillations in main-sequence solar-like stars and subgiants \cite{Cha10}. During the first year of operations, the working group 1 of the Kepler Asteroseismic Science Consortium (KASC \url{http://astro.phys.au.dk/KASC/}) observed more than 2000 solar-like stars looking for stellar pulsations with 1-month long time series reporting their existence in more than 550 stars \cite{Cha11sci}. From the second year on, around 60 stars have been continuously observed for a year and additional $\sim$ 150 stars have been observed for at least 3 continuous months. All this observational set constitutes an ensemble in which we can perform very accurate asteroseismic analyses. Moreover, stellar evolution studies of solar-like stars are completed by the analysis of several thousands of red giants measured in the {\it Kepler} long-cadence mode (e.g. \cite{Bed10}, \cite{Hub10}). 

\section{Some initial results}
\label{sec:First}

Asteroseismology is a powerful tool to look inside the stars and put strong constraints in structure and evolution models (see e.g. \cite{Met10}, \cite{Kal10}, \cite{Mat11s}). In pulsating stars that host planets, asteroseismology allows us to also place very tight constraints on the exoplanetary systems (e.g. \cite{JCD10}, \cite{Bal11}, \cite{Bat11}).
\begin{figure}[!h]
\sidecaption
\includegraphics[scale=.288, angle=90]{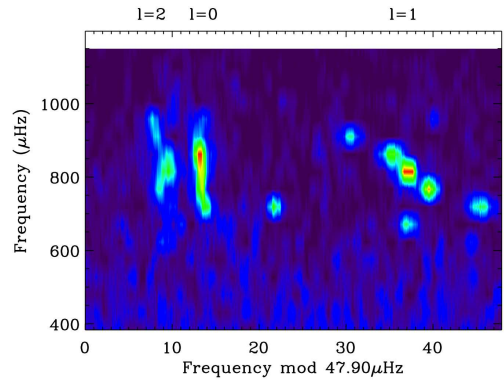}
%
%
\caption{\'Echelle diagram of the post main-sequence star KIC~11395018 showing a mixed mode between the ridges of the l=0 and 1 modes and a few bumped l=1 modes. Adapted from \cite{Mat11b}.}
\label{fig:1}       
\end{figure}

Depending on the signal-to-noise ratio of the modes --which also seems to depend on the magnetic activity of the star \cite{Cha11b}--  on one hand we can have access to only some global seismic parameters (e.g. $\Delta \nu$, $\nu_{\rm{max}}$, $A_{\rm{max}}$ \cite{Ver11}) of the stars using a new generation of automatic procedures that have been set up for solar-like pulsating stars (e.g. \cite{Hek10}, \cite{Hub09},\cite{Mat10}, \cite{Mos09}) including red giants \cite{Hek11}. From these measurements we were able, for example, to verify \cite{Ver11b} the $\log g$ and the radius of the {\it Kepler} Input Catalog \cite{Bro11} inferred from the scaling relations based on solar values (\cite{Kej95} and \cite{Cha11c}). On the other hand --when the signal-to-noise ratio is high enough-- we can characterize very precisely the individual p-mode parameters (e.g. \cite{Mat11b}, \cite{Cam11}). In Fig.~\ref{fig:1} the \'echelle diagram of KIC~11395018 --observed for more than 8 continuous months-- is shown. We clearly identify the ridges of the modes l=2, 0, and 1 from left to right respectively. A nice feature of the power spectrum is the presence of a mixed l=1 mode at 740.29 $\mu$Hz that bumps the modes in the surrounding orders. Indeed mixed modes are very interesting. They behave like acoustic modes near the surface of the stars and as g modes in the interior. They have a very high sensitivity to the core of the stars. In the case of the red giants, many mixed modes can be measured (see e.g. \cite{Bec11}). They can be used to determine the status of the core (if they are already burning He or not \cite{Bed11}) and even to track the internal differential rotation (\cite{Bec11b}, Deheuvels et al. in preparation).  In the case of the subgiants, where only a few  mixed modes are available, their position in frequency depends very much on the age of the star and can be used to further constrain the models \cite{Deh10}.

The high number of stars available to the asteroseismic research allows us to build evolutionary sequences of stars --of similar masses and compositions-- and study the evolution of the physical processes governing their interiors as a function of time,  like a collection of snapshots of the life of a star (see \cite{Sil11} for more details).

With the possible extension of the {\it Kepler mission} to 6 years, the future of the solar-like studies with {\it Kepler} is amazing. In particular, having several years of continuous observations of a several dozens of stars will allow us to study their dynamics: rotation \cite{Gar11b} and magnetic fields. Indeed we will be able to study the surface magnetic activity \cite{Mat10corot} at the same time as the internal one \cite{Gar10}, and the internal structure including the characteristics of the convective zones. The physical processes governing the dynamos will be better constrained and we will improve our knowledge of the solar dynamo, which still hides its secrets (e.g. \cite{Sal09}).
%

\begin{acknowledgement}
RAG wishes to thank the {\it Kepler} team. Funding for this Discovery mission is provided by NASAÕs Science Mission Directorate.  This work has received funding from the European CommunityÕs Seventh Framework Program (FP7/2007-2013) under grant agreement No. 269194 and the NSF under Grant No. NSF PHY05-51164.
\end{acknowledgement}

\end{document}